\documentclass[aps,prl,twocolumn,groupedaddress,superscriptaddress]{revtex4-1}
\usepackage[pdftex]{graphicx}% Include figure files
\usepackage{amsmath}
\usepackage[usenames,dvipsnames]{color}
\usepackage[amssymb]{SIunits}
\usepackage{amssymb}
\newcommand{\ud}[1]{#1^{\dagger}} 
\newcommand{\bra}[1]{\left\langle #1\right|}
\newcommand{\ket}[1]{\left| #1\right\rangle}

\renewcommand{\BibitemShut}{}

\begin{document}

% Use the \preprint command to place your local institutional report
% number in the upper righthand corner of the title page in preprint mode.
% Multiple \preprint commands are allowed.
% Use the 'preprintnumbers' class option to override journal defaults
% to display numbers if necessary
%\preprint{}

%Title of paper
\title{Generation of two identical photons from a quantum dot in a
  microcavity}

% repeat the \author .. \affiliation  etc. as needed
% \email, \thanks, \homepage, \altaffiliation all apply to the current
% author. Explanatory text should go in the []'s, actual e-mail
% address or url should go in the {}'s for \email and \homepage.
% Please use the appropriate macro foreach each type of information

% \affiliation command applies to all authors since the last
% \affiliation command. The \affiliation command should follow the
% other information
% \affiliation can be followed by \email, \homepage, \thanks as well.
\author{E. del Valle} 
\affiliation{Physikdepartment, TU M\"unchen, James-Franck-Str. 1, 85748 Garching, Germany}
\author{A. Gonzalez-Tudela} 
\affiliation{F\'isica Te\'orica de la Materia Condensada, Universidad Aut\'onoma de Madrid, 28049, Madrid, Spain}
\author{E. Cancellieri} 
\affiliation{F\'isica Te\'orica de la Materia Condensada, Universidad Aut\'onoma de Madrid, 28049, Madrid, Spain}
\author{F. P. Laussy} 
\affiliation{Walter Schottky Institut, Technische Universit\"at M\"unchen, Am Coulombwall 3, 85748 Garching, Germany}
\author{C. Tejedor} 
\affiliation{F\'isica Te\'orica de la Materia Condensada, Universidad Aut\'onoma de Madrid, 28049, Madrid, Spain}

\email{elena.delvalle.reboul@gmail.com}

%\thanks{}
%\altaffiliation{}

%Collaboration name if desired (requires use of superscriptaddress
%option in \documentclass). \noaffiliation is required (may also be
%used with the \author command).
%\collaboration can be followed by \email, \homepage, \thanks as well.
%\collaboration{}
%\noaffiliation

\date{\today}

\begin{abstract}
  We propose and characterize a two-photon emitter in a highly
  polarised, monochromatic and directional beam, realized by means of
  a quantum dot embedded in a linearly polarized cavity. In our
  scheme, the cavity frequency is tuned to half the frequency of the
  biexciton (two excitons with opposite spins) and largely detuned
  from the excitons thanks to the large biexciton binding energy.  We
  show how the emission can be Purcell enhanced by several orders of
  magnitude into the two-photon channel for available experimental
  systems.
\end{abstract}

% insert suggested PACS numbers in braces on next line
% insert suggested keywords - APS authors don't need to do this
%\keywords{}

%\maketitle must follow title, authors, abstract, \pacs, and \keywords
\maketitle

Sources of pairs of identical photons are fundamental devices in
quantum metrology~\cite{nagata2007}, quantum communication and
cryptography~\cite{simon2007,cirac1997,duan2001}, linear-optics
quantum computation~\cite{kok2007,lanyon2008}, and even for
fundamental tests of quantum mechanics like hidden variables
interpretations~\cite{aspect1982,collins2002}. A number of devices
have been proposed and experimentally demonstrated with atomic gases
\cite{thompson2006} or nonlinear crystals \cite{nagata2007}. The
realization of such devices, however, is a highly nontrivial task
since, in order to be useful, the generated photons need to be almost
identical, extremely narrow-band and be generated with an extremely
high repetition rate.  Some of us and coworkers have recently proposed
a scheme based on a single quantum dot embedded in a
microcavity~\cite{delvalle10a}, which theoretically fulfils all the
above requirements and, moreover, is particularly promising for
scalable technological implementations.  The principle relies on the
biexciton (the occupation of the quantum dot by two excitons of
opposite spins) being brought in resonance with twice the cavity
photon energy. Thanks to the large biexciton binding energy,
single-photon processes are detuned and are thus effectively
suppressed, while simultaneous two-photon emission is Purcell
enhanced. This effect has been recently demonstrated
experimentally~\cite{arXiv_ota11a}. In the experiment, as in the
initial proposal~\cite{delvalle10a}, the signature for the two-photon
emission is a strong emission enhancement of the cavity mode when
hitting the biexciton two-photon resonance. Because of incoherent
excitation used in both the theoretical proposal and its experimental
realization, the quantum character of the two-photon emission is not
directly demonstrated nor quantified~\footnote{The Authors of
  Ref.~\cite{arXiv_ota11a} also realize this limitation and speculate
  on the scheme that we analyze here in details.}. Here, we upgrade to
a configuration that is nowadays experimentally accessible, where the
quantum dot is initially prepared in a pure biexciton
state~\cite{dousse2010,stufler2006,flissikowski2004}, and analyze in
details the underlying microscopic mechanisms, demonstrating the
perfect two-photon character of the emission beyond a mere enhancement
at the expected energy. We show how the two-photon state is created by
the system in a chain of virtual processes that cannot be broken apart
in physical one-photon states. Our understanding is analytical and
allows for optimisation of a practical setup, enabling the realization
of a practical source of two simultaneous and indistinguishable
photons in a monolithic semiconductor device.

\begin{figure}[t]
  \centering
  \includegraphics[width=\linewidth]{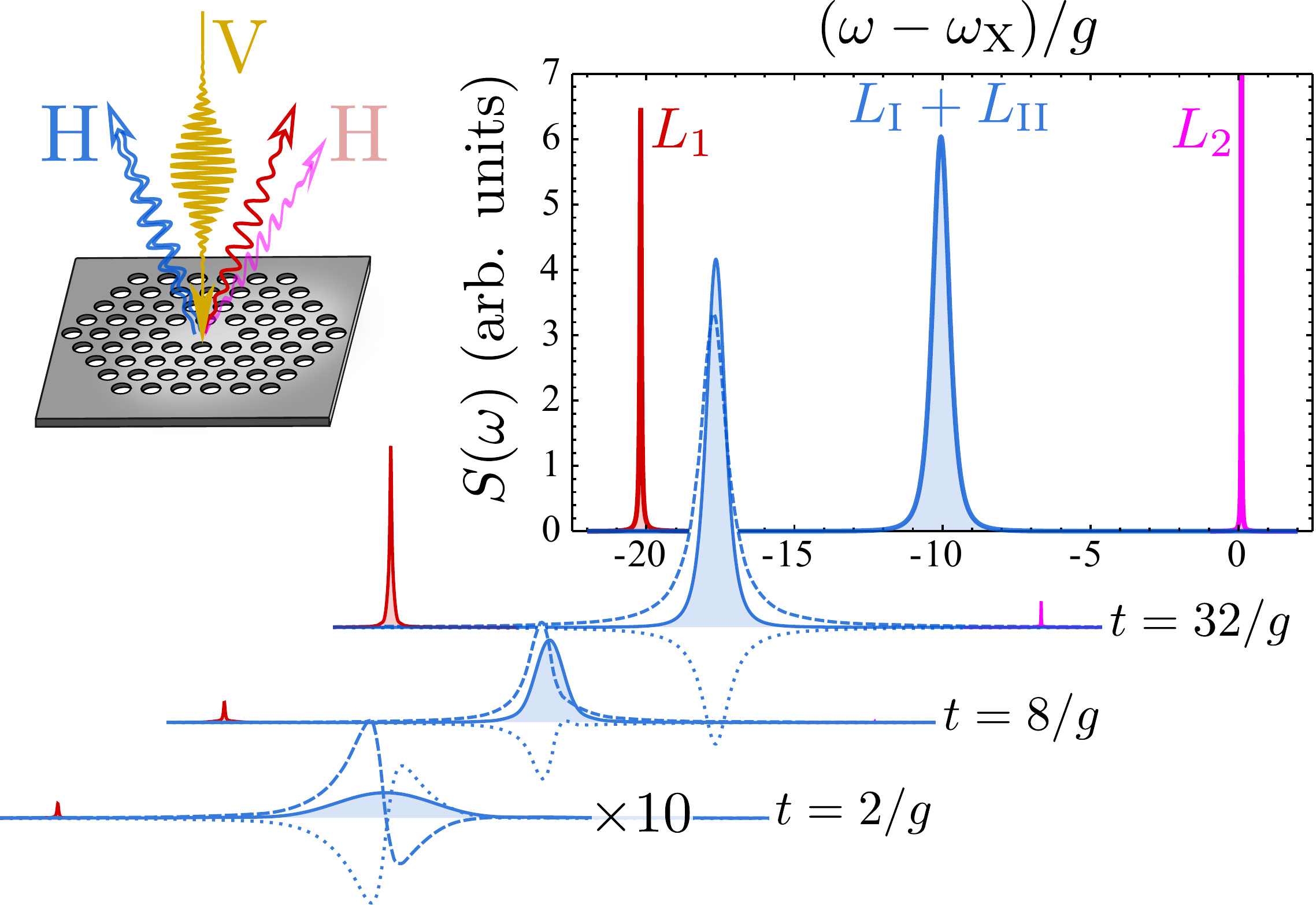}
  \caption{(color online) Cavity spectra of emission $S(t,\omega)$ at
    the two-photon resonance for different times (unframed) and
    integrated over all times (framed).  They feature the 2P peak at
    $\omega_a\approx-\chi/2$ (central, blue) and the two 1P peaks at
    $\omega_1\approx-\chi$ (left, red) and $\omega_2\approx0$ (right,
    pink). The 2P peak cannot be decomposed into two physical
    processes. Parameters: $\chi=20g$, $\kappa=g$,
    $\gamma=5\times10^{-4}g$.}
  \label{fig:1}
\end{figure}

The characteristic spectral profile of the cavity-assisted two-photon
emission is shown in Fig.~\ref{fig:1}, with a central peak that is
strongly enhanced at the two-photon resonance, corroborating its
two-photon character, and surrounded by standard (single-photon)
de-excitation~\cite{delvalle10a,arXiv_ota11a}. The photon-pair peak is
spectrally narrow and isolated from the other events, that can never
be completely avoided, so the source is appealing on practical
grounds. The Hamiltonian of the system reads~\cite{delvalle10a}:
\begin{multline}
  \label{eq:ThuApr14002810CEST2011}
  H(t)=(\omega_a-\omega_\mathrm{L})\ud{a}a+\sum_{i=\uparrow,\downarrow}(\omega_\mathrm{X}-\omega_\mathrm{L})\ud{\sigma}_i\sigma_i-\chi
  \ud{\sigma}_\uparrow\sigma_\uparrow\ud{\sigma}_\downarrow\sigma_\downarrow
  \\+\sum_{i=\uparrow,\downarrow}\Big[g_i(\ud{a}\sigma_i+a\ud{\sigma}_i)+\Omega_i(t)(\sigma_i+\ud{\sigma}_i)\Big]\,,
\end{multline}
where we have included $i=\uparrow,\downarrow$ the spin-up and
spin-down degrees of freedom for the excitonic states $\sigma_i$
(fermions) with common frequency $\omega_\mathrm{X}$ and $a$ the
cavity field annihilation operator (boson) with frequency~$\omega_a$.
The cavity mode can have a strong polarization, say linearly polarized
in the horizontal direction for a photonic crystal, a case we shall
assume in the following. The biexciton binding energy~$\chi$ allows to
bring the biexciton energy $\omega_\mathrm{B}$ in resonance with the
two-photon energy while detuning all other excitonic emissions from
the cavity mode. It is red (blue) shifted if the biexciton is
``bound'' (``antibound''), giving rise to a positive $\chi>0$
(negative $\chi<0$) binding energy
$\chi=2\omega_\mathrm{X}-\omega_\mathrm{\mathrm{B}}$. Our scheme works
with both the bound and antibound biexciton. Without loss of
generality, we assume $\chi>0$, with the added advantage of being less
affected by pure dephasing and coupling to phonons, that we
neglect~\cite{machnikowski08a}.  This binding energy is typically
large ($\chi\approx\unit{400}{\micro\electronvolt}$) as compared to
splittings between excitonic states
($\approx\unit{10}{\micro\electronvolt}$)~\cite{arXiv_ota11a}, which
is ideal for our purpose. We will assume an equal coupling of both
excitons to the linearly polarized mode of the cavity,
$g_\uparrow=g_\downarrow=g/\sqrt{2}$, and take $g$ as the unit in the
remaining of the text. The Hilbert space of the quantum dot is
spanned, in its natural basis of circularly polarised states, by the
ground $\ket{\mathrm{G}}$, spin-up $\ket{\uparrow}$, spin-down
$\ket{\downarrow}$ and biexciton $\ket{\mathrm{B}}$ states. In the
linearly polarised basis, the excitonic states are
$\ket{\mathrm{H}}=(\ket{\uparrow}+\ket{\downarrow})/\sqrt{2}$ and
$\ket{\mathrm{V}}=(\ket{\uparrow}-\ket{\downarrow})/\sqrt{2}$. The
dot-cavity joint Hilbert space includes the photonic number $n$:
$\ket{j,n}$, where $j=\mathrm{G}$, $\mathrm{V}$, $\mathrm{H}$ and
$\mathrm{B}$, with $n\in\mathbb{N}$.

\begin{figure}[t]
  \centering
  \includegraphics[width=\linewidth]{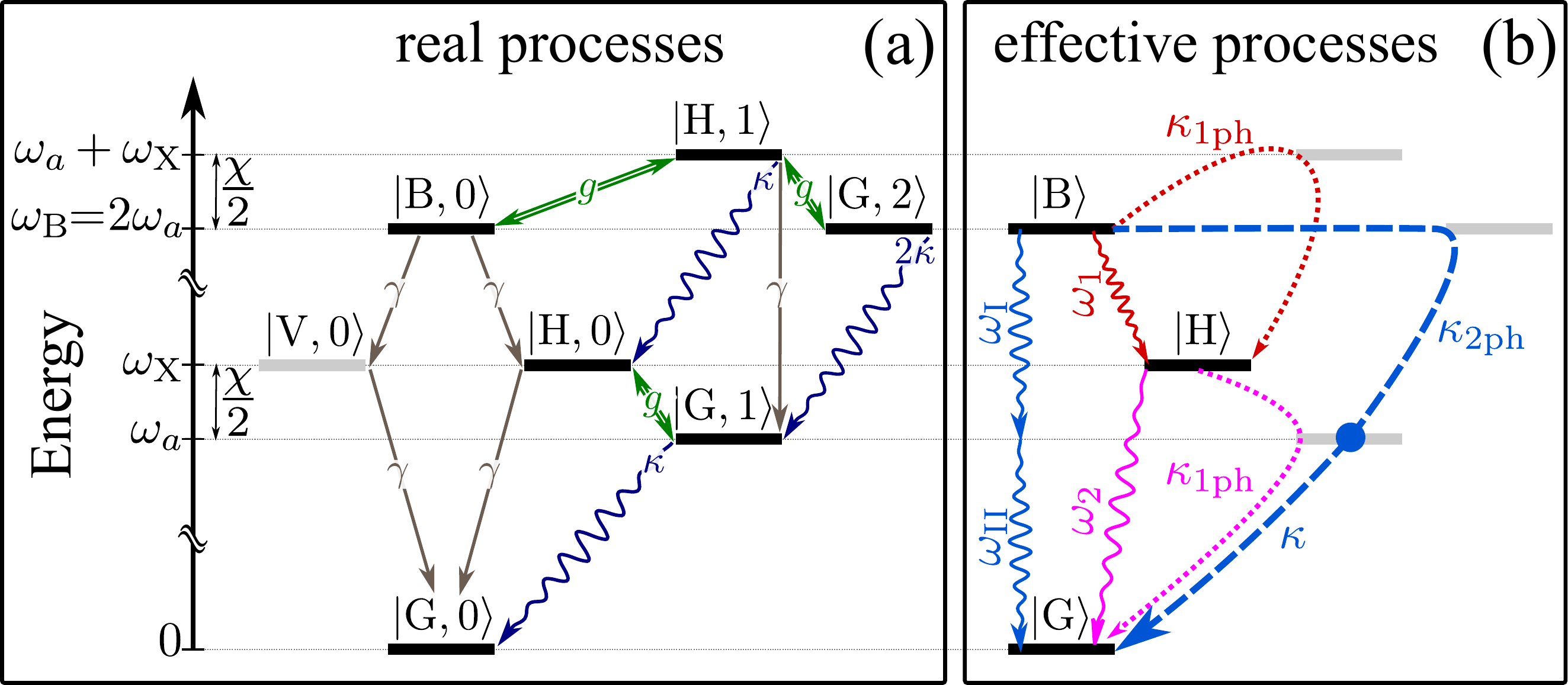}
  \caption{(color online) Level scheme of a quantum dot coupled to a
    cavity mode with linear polarization~H at the 2PR. In (a), the
    microscopic configuration and in (b) the effective processes
    taking place in the de-excitation of the biexciton. Those
    involving the cavity are, on the one hand, through the emission of
    two real and distinguishable photons $\omega_1$ and~$\omega_2$ (in
    dotted red and pink), and, on the other hand, through the
    simultaneous emission of one two-photon state at $\omega_a$ (in
    dashed blue), labelled $\omega_\mathrm{I,II}$.}
  \label{fig:2}
\end{figure}

The quantum dot is excited by a laser of amplitude $\Omega_i(t)$ and
frequency~$\omega_\mathrm{L}$, that brings it in the biexciton state
through two-photon absorption. This can be realised via an appropriate
pulse or sequence of pulses. The laser polarization should be taken
orthogonal to that of the cavity,
$\Omega_\uparrow(t)=-\Omega_\downarrow(t)=\Omega(t)/\sqrt{2}$, so that
the latter is not affected by the excitation process. Coherent control
of the biexciton has been reported in several
works~\cite{dousse2010,stufler2006,flissikowski2004} and we will
assume the biexciton in an empty cavity, $\ket{\mathrm{B},0}$, as the
initial state following the pulse. The laser frequency should be set
to match the two-photon resonance,
$\omega_\mathrm{L}=\omega_\mathrm{B}/2=\omega_\mathrm{X}-\chi/2$. 
With the previous considerations, the Hamiltonian in the basis of
linearly polarized states reads:
\begin{multline}
  \label{eq:SatJun4140529CEST2011}
  H=\omega_a\ud{a}a+\omega_\mathrm{X}
  (\ket{\mathrm{H}}\bra{\mathrm{H}}+\ket{\mathrm{V}}\bra{\mathrm{V}})+(2\omega_\mathrm{X}-\chi)\ket{\mathrm{B}}\bra{\mathrm{B}}\\
  +g \Big[ \ud{a} (\ket{\mathrm{G}}\bra{\mathrm{H}}+\ket{\mathrm{H}}\bra{\mathrm{B}})+\mathrm{h.c.}\Big]\,,
\end{multline}
where it now appears explicitly that the cavity couples only to its
corresponding linear polarization ($\mathrm{H}$). Dissipation affects
the bare states, i.e., in the spin-up/spin-down basis, yielding a
master equation:
\begin{equation}
  \label{eq:ThuApr14005647CEST2011}
  \partial_t\rho=i[\rho,H]
  +\frac{\kappa}{2}\mathcal{L}_a(\rho)+\frac{\gamma}{2}\sum_{i=\uparrow,\downarrow}\Big[\mathcal{L}_{\ket{\mathrm{G}}\bra{i}}+\mathcal{L}_{\ket{i}\bra{\mathrm{B}}}\Big](\rho)
  \,,
\end{equation}
where~$\mathcal{L}_c(\rho)=2c\rho\ud{c}-\ud{c}c\rho-\rho \ud{c}c$,
with $\kappa$ the cavity losses and $\gamma$ the exciton relaxation
rates. Fig.~\ref{fig:2} shows the configuration of levels involved in
the biexciton de-excitation, that is truncated self-consistently. The
coherent coupling ($g$) is represented by bidirectional (green)
arrows, spontaneous decay ($\gamma$) by straight (gray) arrows and
cavity decay ($\kappa$) by curly (blue) arrows, each of them linking
in a reversible ($g$) or irreversible ($\gamma$, $\kappa$) way the
different levels.

A \emph{one-photon resonance} (1PR) is realized when the cavity is set
at resonance with one of the excitonic transitions:
$\ket{\mathrm{B},0}\rightarrow \ket{\mathrm{H},0}$ with frequency
$\omega_1\approx\omega_\mathrm{B}-\omega_\mathrm{X}$ or
$\ket{\mathrm{H},0}\rightarrow\ket{\mathrm{G},0}$ with frequency
$\omega_2\approx\omega_\mathrm{X}$. The resonant single-photon
emission is then enhanced into the cavity mode according to the
conventional scenario~\cite{gerard98a}, with a Purcell decay rate
$\gamma_\mathrm{P}=4g^2/\kappa$.  A \emph{two-photon resonance} (2PR)
is realized when the transition $\ket{\mathrm{B},0}\rightarrow
\ket{\mathrm{G},0}$ matches energetically the emission of two cavity
photons~\cite{delvalle10a}:
\begin{equation}
  \label{eq:SatJun4142833CEST2011}
  \omega_a\approx \omega_\mathrm{X}-\chi/2 \quad \mathrm{with}\quad \chi\gg g\,,\kappa\,,\gamma\,.
\end{equation}
This process also benefits from Purcell enhancement. In fact, if the
decay rates~$\gamma$ and~$\kappa$ are small enough, two-photon Rabi
oscillations between states $\ket{\mathrm{B},0}$ and
$\ket{\mathrm{G},2}$ are even realized, with a characteristic
frequency~$g_\mathrm{2P}\approx
4g^2/(\sqrt{2}\chi)$~\cite{delvalle10a}. Note that in
Eq.~(\ref{eq:SatJun4142833CEST2011}), we have neglected the small
Stark shifts~$\sim g_\mathrm{2P}$, which should be taken into account
to achieve maximum Rabi amplitude.  In this text, to remain within
experimentally achievable configurations, we consider systems in
strong coupling, $g\gtrapprox\kappa$, but not so much that the
two-photon oscillations actually take place, that is, we remain within
the 2P weakly coupled regime, $4g_\mathrm{2P} \ll \kappa$. The
one-photon Rabi oscillations (e.g.,
$\ket{\mathrm{B},0}\leftrightarrow\ket{\mathrm{H},1}$) still take
place at the frequency $g$ but, as they are largely detuned, the
coupling strength effectively reduces to $g_\mathrm{1P}\approx
g/\sqrt{1+[\chi/(\gamma+\kappa)]^2}\approx g
\kappa/\chi$~\cite{laussy09a}.

To characterize and analyze the main output of the system, shown in
Fig.~\ref{fig:1}, we study the time-resolved power spectrum
$S(t,\omega)%\propto \iint_{-\infty}^{t} dt_1 dt_2 e^{i\omega(t-t_1)}e^{-i\omega(t-t_2)}\langle a^{\dagger} (t_1) a (t_2)\rangle\\
\propto \Re\int_0^{t}dT \int_0^{t-T}d\tau e^{i\omega\tau}\langle
a^{\dagger}(T)a(T+\tau)\rangle$~\cite{eberly77a} that we compute as:
\begin{multline}
  \label{eq:ThuApr14014048CEST2011}  
  S(t,\omega)=\frac1\pi\sum_{\alpha\in\{1,2,\mathrm{I},\mathrm{II},\dots\}}\frac{L_\alpha(\gamma_\alpha/2)-K_\alpha(\omega-\omega_\alpha)}{(\gamma_\alpha/2)^2+(\omega-\omega_\alpha)^2}\,,
\end{multline}
where we emphasised in the sum four dominant processes labelled $1$,
$2$, $\mathrm{I}$ and~$\mathrm{II}$ (results below include all
processes). Each $\alpha$ corresponds to a transition in the system,
characterised by its frequency ($\omega_\alpha$) and broadening
($\gamma_\alpha$) on the one hand, which allow us to identify its
microscopic origin, as discussed below, and its intensity $L_\alpha$
and interferences with other transitions $K_\alpha$~\cite{laussy09a}
on the other hand. The time dependent spectra of emission can be
measured experimentally with a streak camera~\cite{wiersig09a}.

There are two channels of de-excitation: via the cavity mode (through
the annihilation of a photon $a$) or via spontaneous emission into the
leaky modes (related to the four excitonic lowering operators).  With
the biexciton state in an empty cavity, $\ket{\mathrm{B},0}$, as the
initial condition, we identify three de-excitation mechanisms of the
system.  We now describe them in turns.

$i$) The first decay route is a cascade of two spontaneous emissions,
from $\ket{B}$ to~$\ket{H}$ (or $\ket{V}$) in a first time, and then
from $\ket{H}$ (or $\ket{V}$) to $\ket{G}$ in a second time, as shown
in straight (gray) lines in Fig.~\ref{fig:2}(a).  This decay into
leaky modes is at the excitonic energies, $\omega_1$, $\omega_2$, and
is a direct process with a straightforward microscopic origin as a
transition between two states. Each process happens at the rate
$\gamma$, so that, as far as the biexciton is concerned, its total
rate of de-excitation through this channel is $2\gamma$. The effect of
this channel is to reduce the efficiency of de-excitation through the
cavity mode, which is the one of interest. This can be kept small by
choosing a system with a small $\gamma$.

$ii$) The second decay route is another cascade of one-photon
emissions, but now through the cavity mode, namely from $\ket{B}$ to
$\ket{G}$ passing by~$\ket{H}$. It is shown in dotted lines in
Fig.~\ref{fig:2}(b). It effectively amounts to two consecutive photons
into the cavity mode at the excitonic energies $\omega_1$ and
$\omega_2$, also shown (with the same color code) in
Fig.~\ref{fig:2}(b), but the microscopic origin is now more complex,
as it involves virtual intermediate states. The first photon (1) is
emitted through the process
$\ket{\mathrm{B},0}\xrightarrow{\ket{\mathrm{H},1}}\ket{\mathrm{H},0}$,
via the off-resonant (``virtual'') state $\ket{\mathrm{H},1}$ and the
second (2), similarly through the process
$\ket{\mathrm{H},0}\xrightarrow{\ket{\mathrm{G},1}}\ket{\mathrm{G},0}$. These
transitions occur at the Purcell rate~$\kappa_\mathrm{1P}\approx 4
g_\mathrm{1P}^2/\kappa \approx T_\mathrm{1P}^2 \kappa$, where
$T_\mathrm{1P}=2g/\chi$ is the effective mixing parameter between
states $\ket{\mathrm{B},0}$--$\ket{\mathrm{H},1}$ and
$\ket{\mathrm{H},0}$--$\ket{\mathrm{G},1}$. The positions and
broadenings are more precisely given by
$\omega_1\approx-\chi-2g^2/\chi$, $\omega_2\approx 2g^2/\chi$ and
$\gamma_1\approx 3\gamma$, $\gamma_2\approx \gamma$.

\begin{figure}[t]
  \centering
  \includegraphics[width=.85\linewidth]{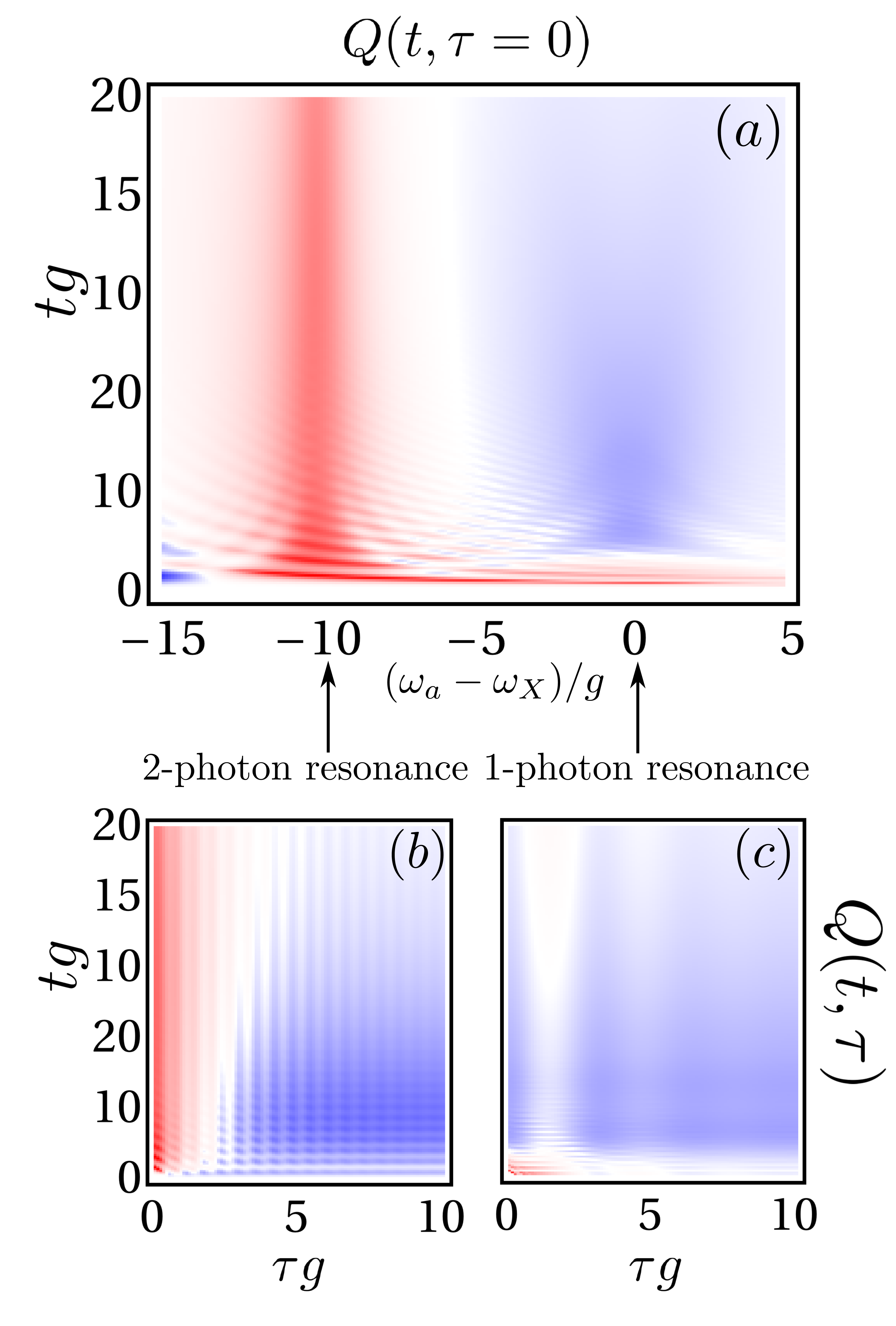}
  \caption{(color online) (a) Mandel parameter $Q(t,\tau=0)$ as a
    function of the cavity frequency~$\omega_a$, for a set of typical
    parameters ($\chi=20g$, $\kappa=0.5g$
    and~$\gamma=5\times10^{-3}g$). $Q(t,\tau)$ is shown below at the
    two relevant resonances, two-photon (a) and one-photon (b).  There
    is a change in the statistics from antibunching $<0$ (1PR),
    colored in blue, to bunching $>0$ (2PR), colored in red.}
  \label{fig:3}
\end{figure}

\begin{figure}[t]
  \centering
  \includegraphics[width=.75\linewidth]{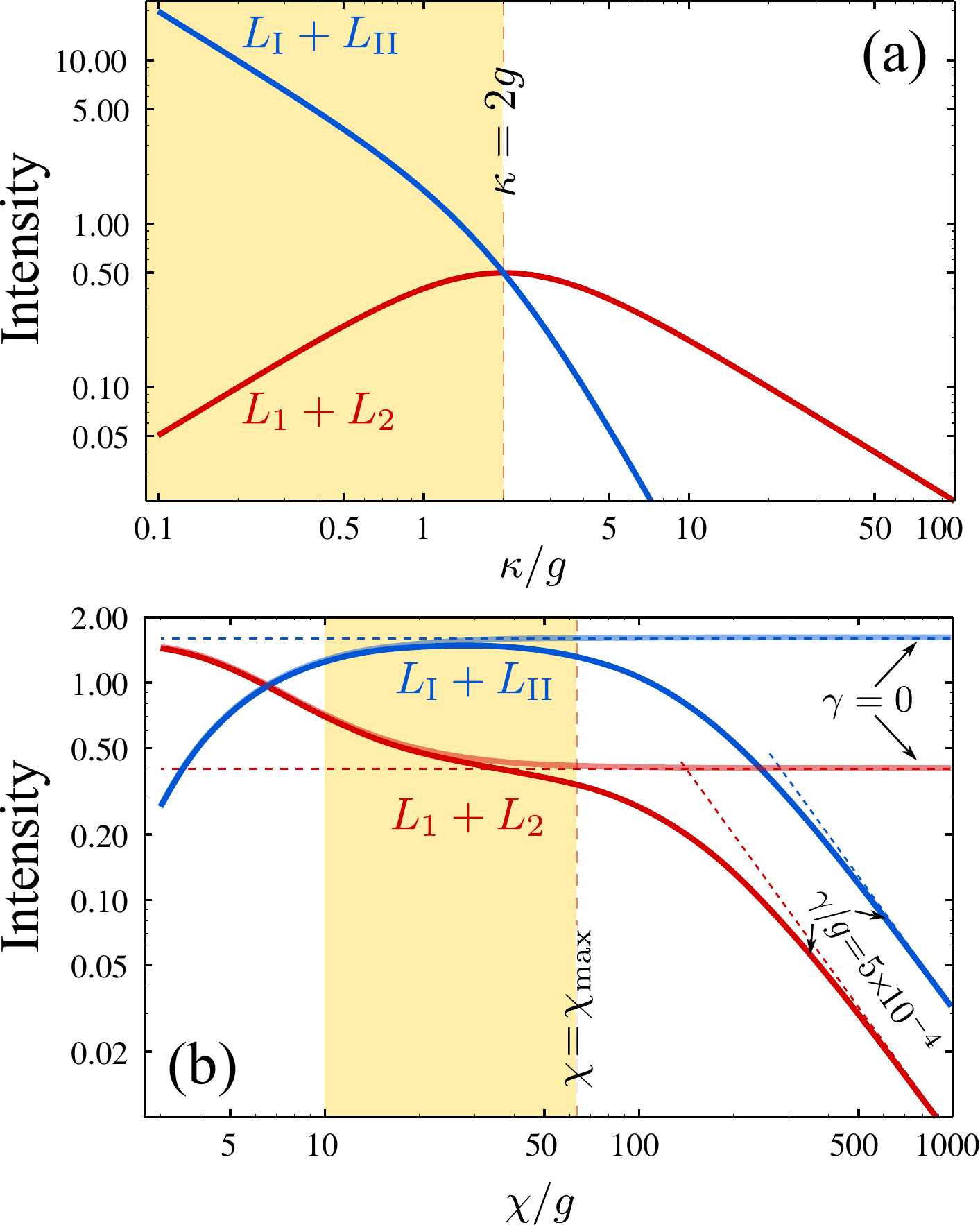}
  \caption{(color online) (a) Intensity $L_\mathrm{I}+L_\mathrm{II}$
    of emission in the 2P channel (blue) and $L_1+L_2$ in the two 1P
    channels (red) as a function of $\kappa$ in the ideal case of
    nonradiative emission, $\gamma=0$ and~$\chi\rightarrow\infty$. The
    shaded (yellow) area $\kappa<2g$ shows the region where the 2P
    emission dominates. (b) Same as above for $\kappa=g$ as a function
    of $\chi$, that must be large enough so that 1P are suppressed and
    small enough to maintain a high cavity emission efficiency in the
    realistic case of nonzero~$\gamma$.}
  \label{fig:4}
\end{figure}

$iii$) Finally, the central event in our proposal is formed by the
third channel of de-excitation of the biexciton, namely, the emission
into the cavity mode of two simultaneous and indistinguishable photons
with a frequency very close to that of the
cavity~$\omega_\mathrm{I}\approx\omega_\mathrm{II}\approx\omega_a$. This
process is sketched by the single dashed (blue) line in
Fig.~\ref{fig:2}(b), with an intermediate step marked by a point at
$\ket{\mathrm{G},1}$. Effectively, this amounts to the generation of a
two-photon state, represented by the two curly
transitions~$\omega_{\mathrm{I},\mathrm{II}}$ in
Fig.~\ref{fig:2}(b). The two indices I and~II strictly correspond to
transitions that arise in the spectral
decomposition~(\ref{eq:ThuApr14014048CEST2011}), namely,
$\ket{\mathrm{B},0}\xrightarrow{\ket{\mathrm{G},2}}\ket{\mathrm{G},1}$
for the first sequence of events, I, and the closing of the path,
$\ket{\mathrm{G},1}\rightarrow\ket{\mathrm{G},0}$, for the second
transition, II. Although we have used I and~II in Fig.~\ref{fig:2} to
label the two photons for the sake of illustration, these two photons
are indistinguishable and cannot be interpreted as real events taken
in isolation in association with the above sequences of
transitions. Indeed, each event gives rise to an unphysical spectrum
(assuming negative values) and only when both processes are taken
together, they interfere to sum to a physical spectrum which can be
interpreted as a probability of (two-photon) detection. This
decomposition of the two-photon (central) peak is shown in
Fig.~\ref{fig:1} in the time-dependent spectra, with the process I
shown in a dotted line and~II in dashed line. They sum to the physical
(observable) peak, in solid line. Both peaks grow together in time and
develop an asymmetry, one (I) being completely positive, the other
(II) completely negative. None, not even the fully positive peak, can
be observed in isolation. In contrast, the single-photon peaks on both
sides (red and pink), are formed by single, isolated transitions,
showing their real (as opposed to virtual) nature.  The two-photon
emission is enhanced by the Purcell rate~$\kappa_\mathrm{2P}\approx 4
g_\mathrm{2P}^2/(2\kappa)\approx T_\mathrm{2P}^2 2\kappa$, where
$T_\mathrm{2P}=g_\mathrm{2P}/\kappa$ is the effective mixing parameter
between states $\ket{\mathrm{B},0}$-$\ket{\mathrm{G},2}$. We use
$2\kappa$ because this is the decay rate of the intermediate state
$\ket{\mathrm{G},2}$. One transition appears, more precisely, at
$\omega_\mathrm{I}\approx-\chi/2+2g^2/\chi$ with broadening
$\gamma_\mathrm{I}\approx \kappa + 2 \gamma$ (this is the sum of the
decay that initial and final states suffer, $\ket{\mathrm{B},0}$ and
$\ket{\mathrm{G},1}$).  The other transition (II) stems from the
direct process $\ket{\mathrm{G},1}\rightarrow\ket{\mathrm{G},0}$.
This transition appears at
$\omega_\mathrm{II}\approx-\chi/2-2g^2/\chi$ with broadening
$\gamma_\mathrm{II}\approx \kappa$.

Another proof of the two-photon character is given by the
time-dependent spectrum, Fig.~\ref{fig:1}. Whereas the single-photon
events grow in succession---first the $L_1$ peak, that populates the
state~$\ket{\mathrm{H}}$, which subsequently decays to
$\ket{\mathrm{G}}$, forming the $L_2$ peak---the two photon peak
arises from the joint and simultaneous contribution of the I and~II
processes. In fact, one can show that at the 2PR,
$L_\mathrm{I}+L_\mathrm{II}\approx2\langle a^{\dagger2}a^2\rangle$,
linking directly the intensity of the peak with the two-photon
emission probability. This can be brought to the experimental test by
resolving the photon statistics in time, $g^{(2)}(t,\tau)=\langle
a^{\dagger}(t)a^{\dagger}(t+\tau)a(t+\tau)a(t)\rangle/[n_a(t)n_a(t+\tau)]$. We
use the Mandel $Q$-parameter, $Q(t,\tau)=n_a(t)(g^{(2)}(t,\tau)-1)$,
that changes sign (negative for anticorrelations). This is shown in
Fig.~\ref{fig:3}. The main panel, (a), shows a strong and sharp
bunching of the emission when the cavity hits the two-photon resonance
(meaning that photons come together, and in our case, in pairs), while
it is antibunched in other cases (photons coming separately). What is
remarkable of the two photon emission is that it is consistently
bunched at all times: while the system can emit at any time, when it
does, it emits the two photons together. In contrast, the 1PR emission
which is antibunched as expected when the process is isolated, also
has the possibility to be bunched by fortuitous joint emission of two
photons. This is the case when $\omega_a=\omega_2$, the cavity is then
in resonance with the lower transition, that can start only as a
successor of the upper transition resulting in high probability for
two photons detection, but only at very early times, since one photon
is a precursor of the other one in a cascade of two otherwise
distinguishable events. The proof is complete with the autocorrelation
time~$\tau$, shown in panels~(b) and~(c), further demonstrating that
in the 2PR emission, the two photons arrive at zero time delay (the
emission being less likely again at nonzero delay).

Now that we have demonstrated from various points of view the
two-photon character of the central peak, we aim to maximise it as
compared to all other de-excitation channels.  There are three key
parameters to enhance the 2P emission, $\kappa$, $\gamma$ and $\chi$.
The case $\gamma=0$ and~$\chi\rightarrow\infty$ is the ideal
configuration, where all the emission goes through the cavity:
\begin{equation}
  \label{eq:SatJul2013152CEST2011}
  I_a=\int_0^\infty\langle\ud{a}a\rangle(t)\,dt = 2/\kappa\,,
\end{equation}
which is redistributed between the two possible decay paths as:
\begin{subequations}
  \label{eq:SatJul2013551CEST2011}
  \begin{align}
    &L_1+L_2\approx\frac{\kappa_\mathrm{1P}}{\kappa_\mathrm{1P}+\kappa_\mathrm{2P}}
    I_a \approx \frac{2}{\gamma_\mathrm{P}+\kappa}\,,\\
    &L_\mathrm{I}+L_\mathrm{II}\approx\frac{\kappa_\mathrm{2P}}{\kappa_\mathrm{1P}+\kappa_\mathrm{2P}}
    I_a \approx \frac{2\gamma_\mathrm{P}/\kappa}{\gamma_\mathrm{P}+\kappa}\,.
  \end{align}
\end{subequations}
This is shown in Fig.~\ref{fig:4}(a), where we see that the 2P
emission dominates over the 1P when $\kappa<2g$ (shaded in yellow in
Fig.~\ref{fig:4}(a)), since in this case
$\kappa_\mathrm{2P}>\kappa_\mathrm{1P}$. For cavities with high enough
quality factor (small $\kappa$), the 2P emission is over four orders
of magnitude higher than the 1P, showing that the device is extremely
efficient with favourable technological parameters. 

When $\gamma$ is nonzero, the situation of experimental interest, but
still is the smallest parameter ($\ll \kappa, g\ll\chi$), the channel of
decay it opens leads
to:
\begin{equation}
  \label{eq:SatJul2020709CEST2011}
  I_a =  \int_0^\infty n_a(t)\, dt = \frac{\gamma_\mathrm{P}(\gamma_\mathrm{P}+\kappa)}{\gamma \chi^2}\,,
\end{equation}
which is now redistributed between the two cavity decay paths 
as an increasing function of $\chi^{-2}$:
\begin{subequations}
  \label{eq:SatJul2020704CEST2011}
  \begin{align}
    &L_1+L_2\approx\frac{\kappa_\mathrm{1P}}{\kappa_\mathrm{1P}+\kappa_\mathrm{2P}+2\gamma}
    I_a \approx \frac{\gamma_\mathrm{P} \kappa}{\gamma \chi^2}\,,\\
    &L_\mathrm{I}+L_\mathrm{II}\approx\frac{\kappa_\mathrm{2P}}{\kappa_\mathrm{1P}+\kappa_\mathrm{2P}+2\gamma}
    I_a \approx
    \frac{\gamma_\mathrm{P}^2}{\gamma \chi^2}\,.
  \end{align}
\end{subequations}

This nonzero $\gamma$ case is shown in Fig.~\ref{fig:4}(b), where the
ideal situation can be recovered in a region of $\chi$ bounded by
above by:
\begin{equation}
  \label{eq:SatJul2021809CEST2011}
  \chi_\mathrm{max} = \mathrm{min}(2g\sqrt{\kappa/(2\gamma)}\,,4g^2/\sqrt{2\kappa\gamma})\,,
\end{equation}
that follows from
$2\gamma=\min(\kappa_\mathrm{1P},\kappa_\mathrm{2P})$.  Above
$\chi_\mathrm{max}$, the 2P emission still dominates over 1P emission
but efficiency is spoiled, according to
Eqs.~(\ref{eq:SatJul2020704CEST2011}), that are shown in dashed tilted
lines.

In conclusion, we have presented a scheme where the biexciton is in
two-photon resonance with a microcavity mode, as an efficient
two-photon source, both in terms of the purity of the two-photon state
and of its emission efficiency. The timescale for two-photon emission,
that limits the repetition rate, is of the order of
$\kappa_{2\mathrm{P}}^{-1}$. The quantum character of the two-photon
emission is demonstrated theoretically by a detailed analysis of all
the processes involved in the biexciton de-excitation, which also
allows us to find analytically the optimum conditions for its
realization. We have shown that the two-photons are emitted
simultaneously with no delay in the autocorrelation
time. Experimentally, the ultimate proof of indistinguishability can
be obtained by directing the central peak to a beam-splitter, which
half of the time will separate the photon pair into two ports that can
then be fed in an Hong-Ou-Mandel interferometer.

\begin{acknowledgments}
  We thank F. Troiani, D. Sanvitto, A. Laucht and J. J. Finley for
  discussions. We acknowledge support from the Emmy Noether project HA
  5593/1-1 (DFG), the Marie Curie IEF `SQOD', the Spanish MICINN
  (MAT2008-01555 and CSD2006-00019-QOIT) and CAM
  (S-2009/ESP-1503). A.G.-T. thanks the FPU program (AP2008-00101)
  from the Spanish Ministry of Education.
\end{acknowledgments}

% Create the reference section using BibTeX:
%\bibliographystyle{apsrev}
\bibliography{Sci,books,emi,2phot}
\end{document}